%% file: OBheating.tex
\documentclass[useAMS,usenatbib]{mnras} 
\usepackage{natbib}
\usepackage{graphicx}
\usepackage{setspace}
\usepackage{mathtools}
\usepackage{pdflscape, longtable}
\usepackage{hyperref}

\usepackage[toc,page]{appendix}
\usepackage{multirow}

\newcommand{\HII}{{\sc Hii}}

\newenvironment{jh}{}{}

\def \arcsec {$^{\prime\prime}$}

\title[JCMT GBS: radiative heating.]{ The JCMT Gould Belt Survey: radiative heating by OB stars.}

\author[D. J. Rumble]{D. Rumble$^{1}$, J. Hatchell$^{1}$, H. Kirk$^{2,3}$, K. Pattle$^{4}$\\
$^{1}$Physics and Astronomy, University of Exeter, Stocker Road, Exeter EX4 4QL, UK\\
$^{2}$NRC Herzberg Astronomy and Astrophysics Research Centre, 5071 West Saanich Road, Victoria, BC, V9E 2E7, Canada \\
 $^{3}$Department of Physics and Astronomy, University of Victoria, 3800 Finnerty Road, Victoria, BC, V8P 5C2, Canada\\
 $^{4}$National University of Ireland Galway, University Road, Galway, Ireland H91 TK66}
\begin{document}

\date{}

\maketitle 

\label{firstpage}

\begin{abstract}

Radiative feedback can influence subsequent star formation.  We quantify the heating from OB stars in the local star-forming regions in the  JCMT Gould Belt survey.
Dust temperatures are calculated from 450/850~\micron\ flux ratios from SCUBA-2 observations at the JCMT \begin{jh} assuming a fixed dust opacity spectral index $\beta=1.8$\end{jh}.  Mean dust temperatures are calculated for each submillimetre clump along with projected distances from the main OB star in the region. Temperature vs. distance is fit with a simple model of dust heating by the OB star radiation plus the interstellar radiation field and dust cooling through optically thin radiation.
Classifying the heating sources by spectral type, O-type stars produce the greatest clump average temperature rises and largest heating extent, with temperatures over 40~K and significant heating out to at least 2.4~pc.  Early-type B stars (B4 and above) produce temperatures of over 20~K and significant heating over 0.4~pc.  Late-type B stars show a marginal heating effect \begin{jh}within\end{jh} 0.2~pc.  For a given projected distance, there is a significant scatter in clump temperatures that is due to local heating by other luminous stars in the region, projection effects, or shadowing effects.  
Even in these local,  `low-mass' star-forming regions, radiative feedback is having an effect on parsec scales, with 24\% of the clumps heated to at least 3~K above the 15~K base temperature expected from  heating by only  the interstellar radiation field, and a mean dust temperature for heated clumps of 24~K.

\end{abstract}



\section{Introduction}

Radiative feedback from young, higher-mass stars is thought to affect subsequent star formation by heating the gas and reducing fragmentation \citep{bate:2009uq,Howard:2016aa}.

We have been exploring the observational evidence for this effect in a series of papers looking at heating in local star-forming regions observed as part of the JCMT Gould Belt Survey \citep{gbs}, specifically Perseus~NGC1333 \citep{Hatchell:2013ij}, Serpens~MWC297 \citep{Rumble:2015vn} and the Aquila~W40 complex \citep{Rumble:2016kx}).  Using dust colour temperatures derived from submillimetre ratios, we have found evidence that OB stars heat the dust in nearby filaments and cores and potentially increase their stability against gravitational collapse through raising the Jeans mass.

Although these nearby regions are thought of as low-mass star-forming regions (with the exception of Orion~A), there are O or B stars associated with several of them.  Because only one or two OB stars are associated with each region, it is simpler to consider the influence of those stars on the surrounding cloud material than in high-mass star-forming regions, where the heating effect of several stars is combined.  

In this study, we extend our investigation of radiative feedback to include all clouds in the local star-forming regions within the James Clerk Maxwell Telescope Gould Belt Survey (JCMT GBS), considering the dust heating as a function of stellar spectral type and projected distance from the OB star.

\section{Methods}

\begin{table*}
\centering
\caption{The 30 JCMT GBS (sub)regions included in the \textsc{fellwalker} SCUBA-2 850\,$\micron$ clump catalogue. RA and Dec give the location of the brightest clump in each region. Where a region has a major OB star, its name and classification are listed.}
\label{tbl:regions}
\begin{tabular}{llccccll}
\hline\hline
Subregion        & RA  	& Dec 	&Distance  & Clumps & Clumps                   & Major OB  & Class\\
	          &(J2000) & (J2000) & (pc)               & (total)  &(with $T_{\mathrm d}$)  & star(s) & \\
\hline 
\hline
  
\input{table1.tex}

  \hline
\hline
\end{tabular}
\raggedright

Distance references: a -- \cite{Zucker:2019}; b -- \cite{Zucker:2020}; c -- \cite{Ortiz-Leon:2017ab}; d -- \cite{Ortiz-Leon:2018a}; e -- \cite{Herczeg:2019}; f -- \cite{Ortiz-Leon:2017aa}.  
OB star references:  1 -- \cite{Cutri:2003kx};  2 -- \cite{Guetter:1968ij};  3 -- \cite{Gray:2006bh};  4 -- \cite{Hog:2000kl};  5 -- \cite{Cieza:2007pi};  6 -- \cite{Houk:1988kl};  7 -- \cite{Sota:2011};  8 -- \cite{Abt:1977aa};  9 -- \cite{Strom:1974zr}; 10 -- \cite{Herbst:2008pi}; 11 -- \cite{Walawender:2008zr}; 12 -- \cite{Drew:1997qf}; 13 -- \cite{HoukSwift:1999}; 14 -- \cite{Smith:1985bv}
 
 \end{table*}

\begin{figure*}
\begin{centering}
 \includegraphics[width=\textwidth]{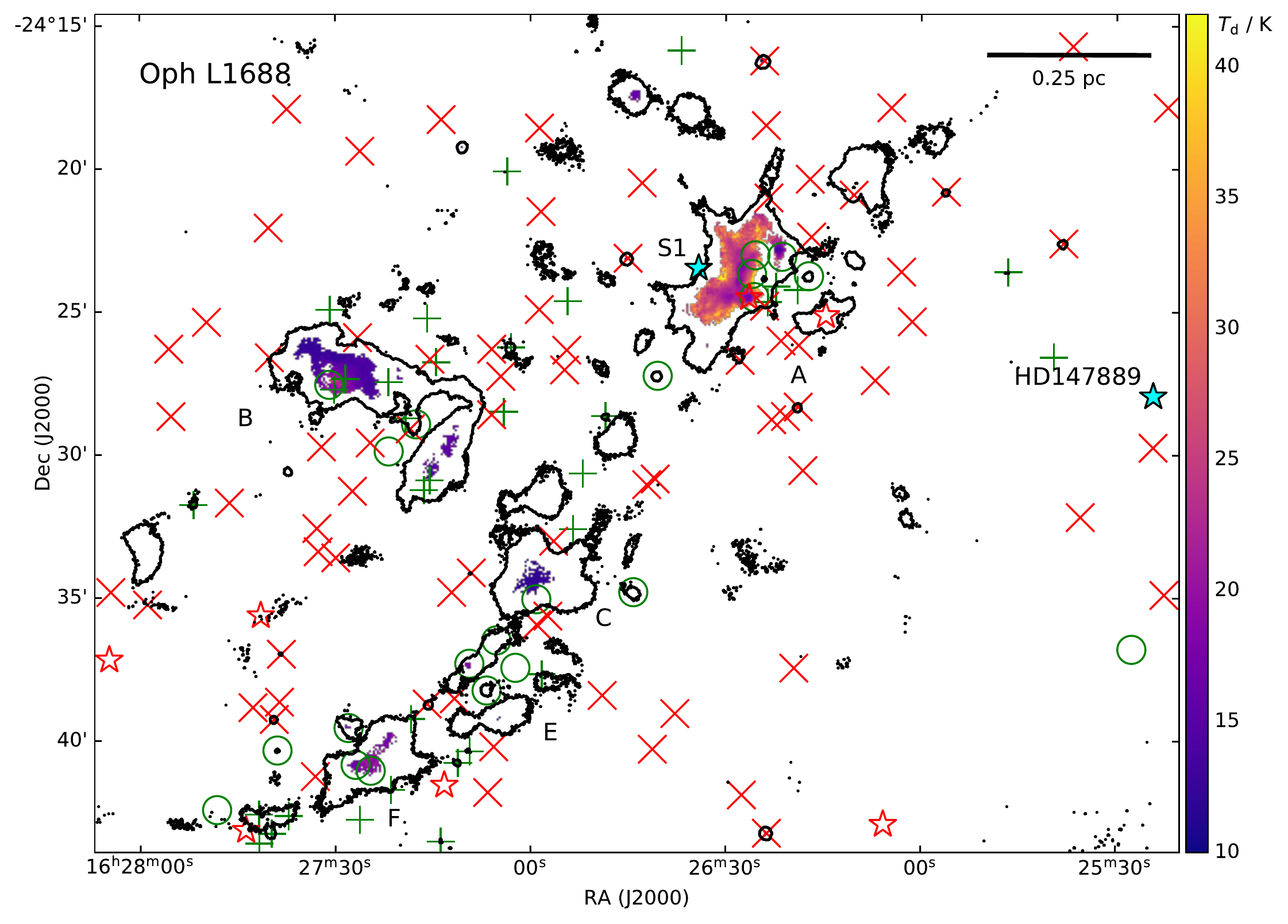}
\caption{SCUBA-2-derived dust colour temperatures in Ophiuchus~L1688 \citep{lynds62}.  The OB stars S1 and HD 147889 are marked as cyan stars.  Also marked are the young stellar object candidates from \citet{Dunham:2015kx} (green circle = Class 0, green plus = Class I, red cross = Class II, red star = Class III).  The 850~\micron\ SNR=3 contour is shown in black (0.26~mJy/arcsec$^2$)  }
\label{fig:Tmap}
\end{centering}
\end{figure*}

Maps of dust colour temperatures based on SCUBA-2 450~\micron\ and 850~\micron\ maps from the  JCMT Gould Belt Survey 
Data Release 1 (DR1) were created for the 30 GBS subregions listed in Table~\ref{tbl:regions}.  For details of the observations (2012 to 2015), data reduction, and maps, see \citet{Kirk:2018} and earlier survey publications \citep{Rumble:2015vn,Pattle:2015ys,Mairs:2015vn,Chun-Yuan-Chen:2016kx,Kirk:2016uq,Rumble:2016kx}.  SCUBA-2 maps contain spatial scales up to $5'$ (0.75~pc at 500~pc) so they trace the dense cores and filaments most directly associated with star formation without the large-scale cloud background \citep{WardThompson:2016aa}.

Dust temperatures $T_{\mathrm d}$ were calculated pixel-by-pixel from the ratio of SCUBA-2 450\,$\micron$ ($S_{\mathrm{450}}$) and 850\,$\micron$ ($S_{\mathrm{850}}$) fluxes following \cite{Reid:2005ly} in using
\begin{equation}
\frac{S_{\mathrm{450}}}{S_{\mathrm{850}}} = \left(\frac{850}{450} \right)^{3+\beta }\left(\frac{\exp(hc/\lambda _{\mathrm{850}}k_{\mathrm{b}}T_{\mathrm{d}})-1}{\exp(hc/\lambda _{\mathrm{450}}k_{\mathrm{b}}T_{\mathrm{d}})-1} \right), \label{eqn:tempb}
\end{equation}
where $\beta$ is the dust opacity spectral index \citep{beckwith:1990fk}.  The 450~\micron\ map was first convolved to match the 850~\micron\ resolution using the \cite{Aniano:2011fk} kernel convolution algorithm adapted for SCUBA-2 images by \cite{Pattle:2015ys} and \cite{Rumble:2016kx} with an analytical model-beam kernel  based on the two component beam model of \cite{Dempsey:2013uq}; see \cite{Rumble:2016kx} for details.  The resulting dust temperature maps have an effective resolution of 14.8\arcsec, comparable to the JCMT 850~\micron\ effective beam FWHM of 14.6\arcsec.  Temperature estimates are limited to the cores and filaments detected at 450~\micron\ signal-to-noise ratios (SNR) greater than 5 { and pixels with temperature uncertainties $<34.1$\% \citep{Rumble:2016kx}.  \begin{jh} This is a choice in favour of reliability over area coverage: relaxing these requirements mostly adds hot, high-uncertainty pixels around the periphery of the clumps.\end{jh}

  An example temperature map for Ophiuchus~L1688 is shown in Fig.~\ref{fig:Tmap}; the full set of dust temperature maps are publicly available from \url{https://dx.doi.org/10.11570/20.0007}

\footnote{A recent reduction of the GBS 450~\micron\ and 850~\micron\ maps is already publicly available at \url{https://doi.org/10.11570/18.0005} \citep{Kirk:2018}}.

  A constant $\beta$ of 1.8 was assumed, consistent with most dust opacities measured in Perseus from joint Herschel/SCUBA-2 fits \citep{Chun-Yuan-Chen:2016kx,Sadavoy:2013qf}, \emph{Planck} observations \citep{Juvela:2015ys} and that used in other GBS papers (\citealt{Hatchell:2005fk}, \citealt{salji:2015fk}, \citealt{Rumble:2015vn} and \citealt{Pattle:2015ys}).  Using a fixed $\beta$ avoids the inherent anticorrelation between $T_{\mathrm d}$ and $\beta$ when fitting fluxes \citep{Chun-Yuan-Chen:2016kx,Shetty:2016gb} at the expense of temperature errors when the assumed $\beta$ is wrong (for example in protostellar discs where $\beta\sim1$ is a better estimate).  \begin{jh}The choice of $\beta=1.8$ is also consistent with the \citet{Ossenkopf:1994vn} dust properties used for modelling in Sect.~\ref{sect:results}.  The absolute dust temperature scale throughout this work depends on the choice of $\beta$ as shown in Fig.~\ref{fig:Tbeta}, with lower $\beta$ values giving higher temperatures for the same observed flux ratio -- for example, the median clump temperature derived in Sect.~\ref{sect:results} would increase to 18.2~K for a lower assumed $\beta=1.6$ or reduce to 12.9~K for $\beta=2.0$, and this spread increases at higher temperatures.  Simultaneous determinations of $\beta$ and temperature are now available for a few local clouds based on joint SCUBA-2 and Herschel analysis \citep{Sadavoy:2013qf,Chun-Yuan-Chen:2016kx,Howard:2019,Howard:2021}; but until these are consistently available for many clouds, this simplifying assumption is necessary.
\end{jh}

\begin{figure}
\begin{centering}
\includegraphics[clip, width=\columnwidth]{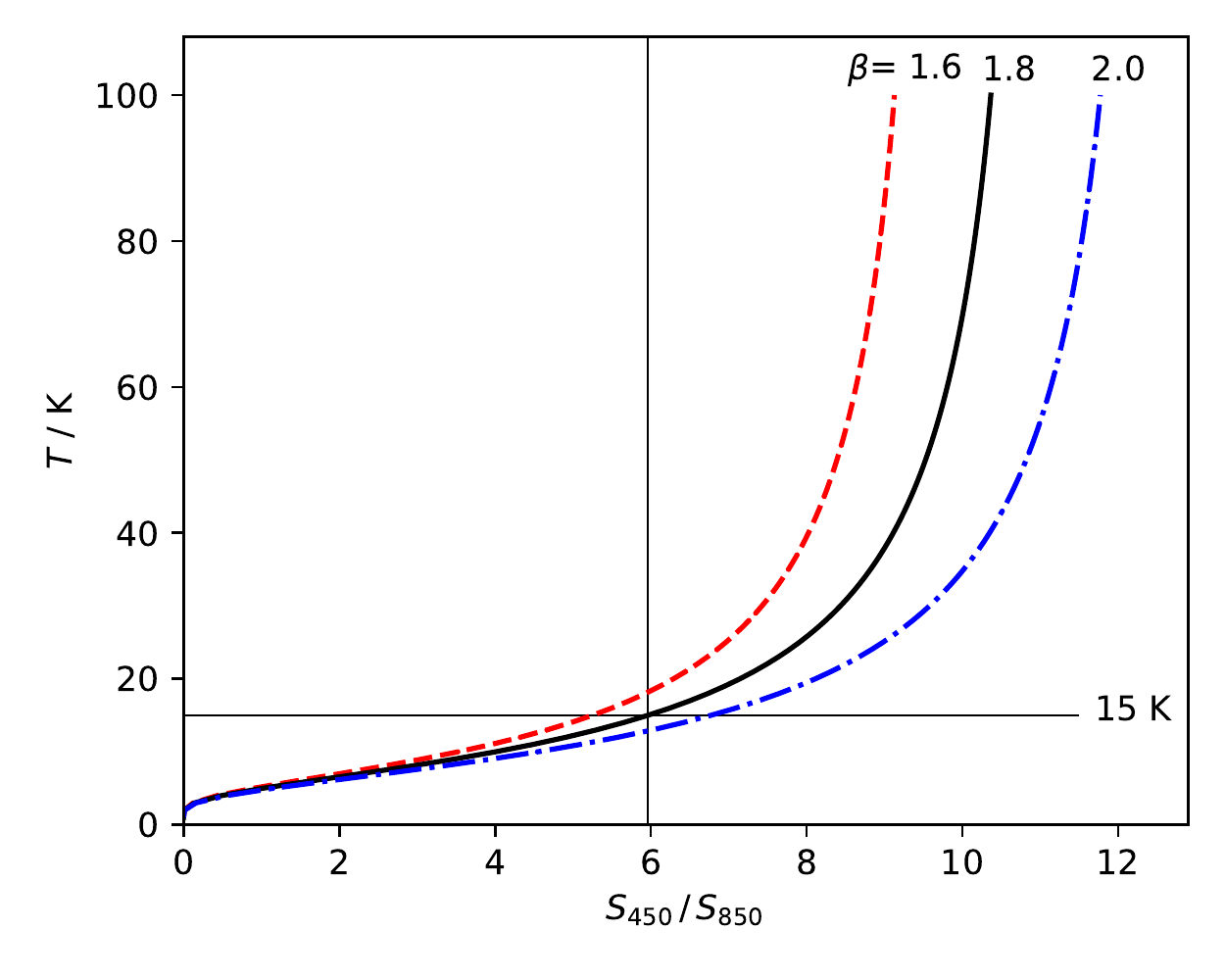}
\caption{Dust temperature as a function of 450\micron / 850\micron\ flux ratio for different dust opacity spectral index $\beta=1.6, 1.8$ (used in this work) and $2.0$.  The ratio $S_{450}/S_{850} =5.96$ corresponding to the median clump temperature of $15.0$~K assuming $\beta=1.8$ is marked for reference.} 
\label{fig:Tbeta}
\end{centering}
\end{figure} 

 The  dust colour temperature obtained by solving Equation~\ref{eqn:tempb} is an average over the line-of-sight components weighted by the emission at 450~\micron\,
 which is longward of the peak of the spectral energy distribution even for dust as cold as 10~K.  In the 450/850 ratio, there is a relatively high contribution from the colder dust  ($<20$~K) that makes up the bulk of the mass of (starless) cores and filaments.    In addition, SCUBA-2 maps spatially filter out scales greater than $5'$ \citep{Mairs:2015,Kirk:2018}, so it is insensitive to extended emission, for example, emission from warm diffuse dust heated by clusters or low-extinction cloud layers.   Nonetheless, emission-weighted line-of-sight average dust temperatures are always higher than the equivalent mass-weighted temperatures. 

 The SCUBA-2 850\,$\micron$ regions were split into clumps using the Starlink \textsc{fellwalker} algorithm \citep{Berry:2015uq} following \cite{Rumble:2016kx} with an 850\,$\micron$ SNR threshold of 5.  Clumps smaller than the beam were rejected, eliminating many Class~II discs (which we wish to exclude as their temperatures are dominated by internal heating and for which our assumed $\beta$ is incorrect) as well as artefacts. In total, 1134 clumps were detected across the 30 subregions listed in Table \ref{tbl:regions} (743 excluding Orion~A).
 Clump sizes along each of the pixel axes were calculated by \textsc{fellwalker} as the root-mean-square flux-weighted offset of each pixel from the clump centroid, equivalent to the Gaussian width for a Gaussian clump, from which we take the geometric mean \citep[for details, see
][]{Berry:2015uq}.  The beam-deconvolved flux weighted clump sizes range between 0.01~pc and 0.37~pc 
with a median of 0.08~pc, larger than a typical protostellar core (0.05~pc, \citealt{Rygl:2013ve}), confirming that the \textsc{fellwalker} clumps represent more extended, filamentary structures.  Average clump temperatures were calculated for the subset of 465/743 clumps for which at least 50\% of the pixels within the clump had a reliable temperature estimate  (i.e. a 450~\micron\ SNR $>5$ and a temperature uncertainty of less than $34.1\%$).

Although it is the best known massive star forming region in the Gould Belt, Orion A is excluded from our analysis for two reasons.  Firstly, temperatures around OMC-1 are so high that the SCUBA-2 450~\micron\ fluxes lie on the Rayleigh-Jeans tail and so Equation \ref{eqn:tempb} cannot be used to calculate reliable temperatures.  Secondly,  there are so many OB stars in Orion that identifying the one(s) heating each clump and assigning a projected distance is not trivial.

The OB (and A) star(s) potentially heating each subregion were compiled based on the \cite{Reed:2003vn} Catalogue of Galactic OB Stars, rejecting line-of-sight contaminants based on proper motions from the \textsc{SIMBAD} database \citep{Wenger:2000ve}.  For regions with multiple OB stars a major OB star was identified (see Table \ref{tbl:regions}): these were selected primarily on spectral class, favouring the most massive stars, but also proximity to any clumps, as in the cases of Ophiuchus~L1688,  Perseus~NGC1333, and NGC2023/24 where lower mass OB stars are directly heating their immediate environment in addition to more distant, higher mass OB stars \citep{Pattle:2015ys,Hatchell:2005fk,Meyer:2008}.  We calculated the projected distance (in parsecs) of each SCUBA-2 clump to its nearest major OB star.

\section{Results and analysis}
\label{sect:results}

\begin{figure*}
\begin{centering}
\includegraphics[clip, width=0.85\textwidth]{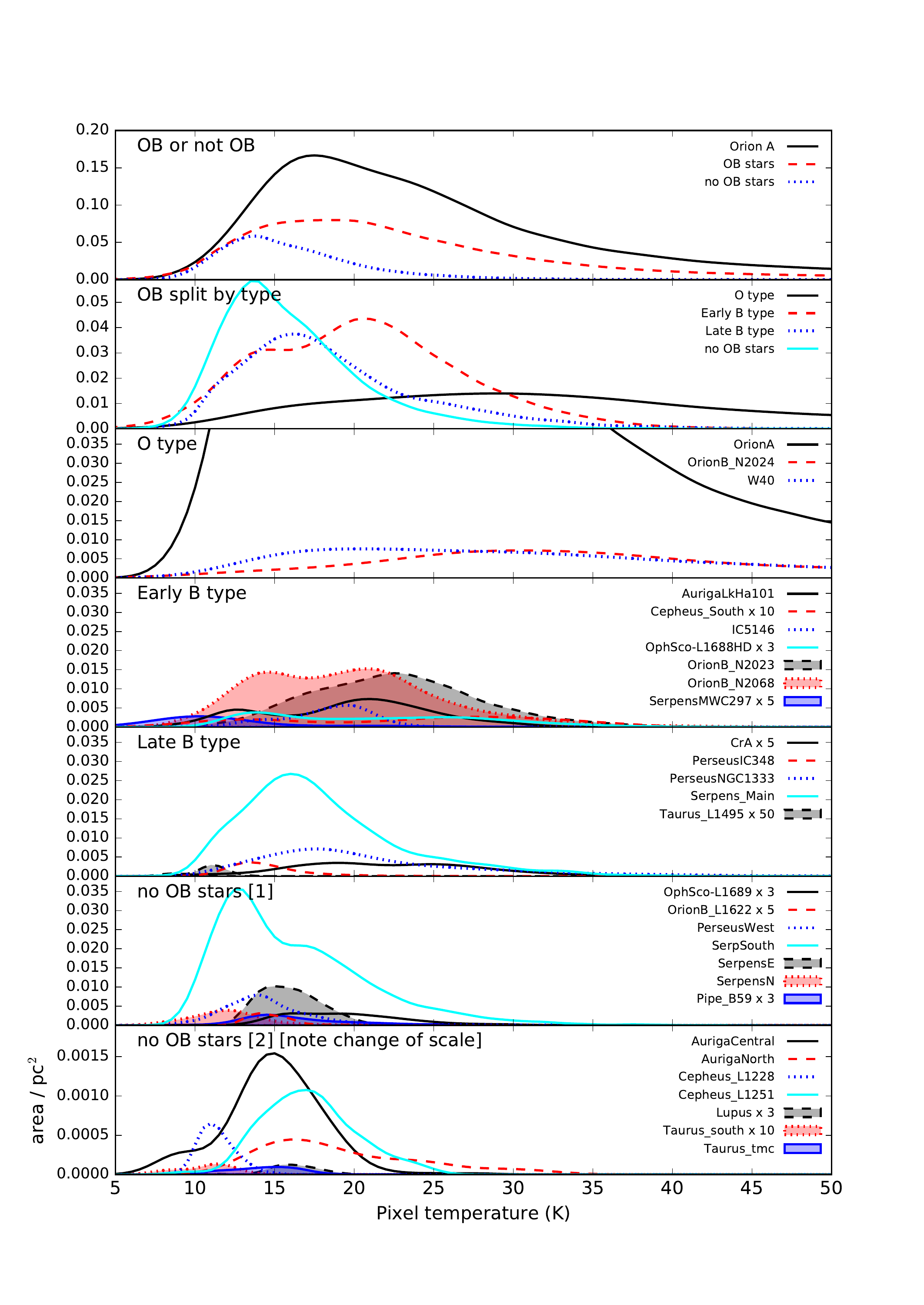} 
\caption{The area of pixels with temperature estimates in each GBS region as a function of temperature.  The probability density function is calculated using a Gaussian Kernel Density Estimator with the \citet{Silverman:1986} rule-of-thumb bandwith estimator and scaled by the pixel area at appropriate distance.  {\bf Top to bottom:}  regions with and without OB stars and (separately) Orion A; regions with OB stars split by spectral type O, early B (B0--4), late B (B5--9 plus A0), no OB stars; 4 panels with distributions for individual regions with O~stars, including Orion A;  early-type B~stars, late-type B~stars, and (two panels) no identified OB stars.  Note changes of scale and that some areas are scaled up as indicated in the legend.  Only pixels with 450/850~\micron\ SNR $> 5$ and temperature uncertainties $<34.1$\% are included so there is a bias towards areas of warm, high column density dust that produce strong emission.   } 
\label{fig:histograms}
\end{centering}
\end{figure*} 

Figure \ref{fig:histograms} shows that the 16 subregions with OB stars contain larger areas of warm dust (20--50~K) than those without OB stars.

In regions with OB stars, the distribution is shifted to higher temperatures with a median temperature of  21~K  compared to 15~K  for regions without OB stars.  Regions with OB stars have a factor 3.3 more pixels above 20~K than regions without OB stars.

We further subdivide the 16 regions with OB stars into three sets --- O-type, Early B-type (B0--B4) and Late B-type (B5--A0) --- based on the classification of the major OB star in the region  (see Table \ref{tbl:regions} for assignments). A0 is included in the Late B class in order to assess the influence of V892Tau in Taurus~L1495.    The distributions of pixel temperatures for each subset, and for the regions individually, are shown in Fig.~\ref{fig:histograms} (second panel from top and lower panels).   Median temperatures are 33~K, 21~K, 18~K  and 15~K for regions with O, early B, late B-type, and no OB stars, respectively.  
Regions with O, early B, or late B stars show greater fractions of warm dust (20--50~K) than regions with no OB stars, with the greatest fraction in the two regions with O stars (Orion~B NGC2024 and W40).

\begin{figure*}
\begin{centering}
  \includegraphics[clip,width=\textwidth]{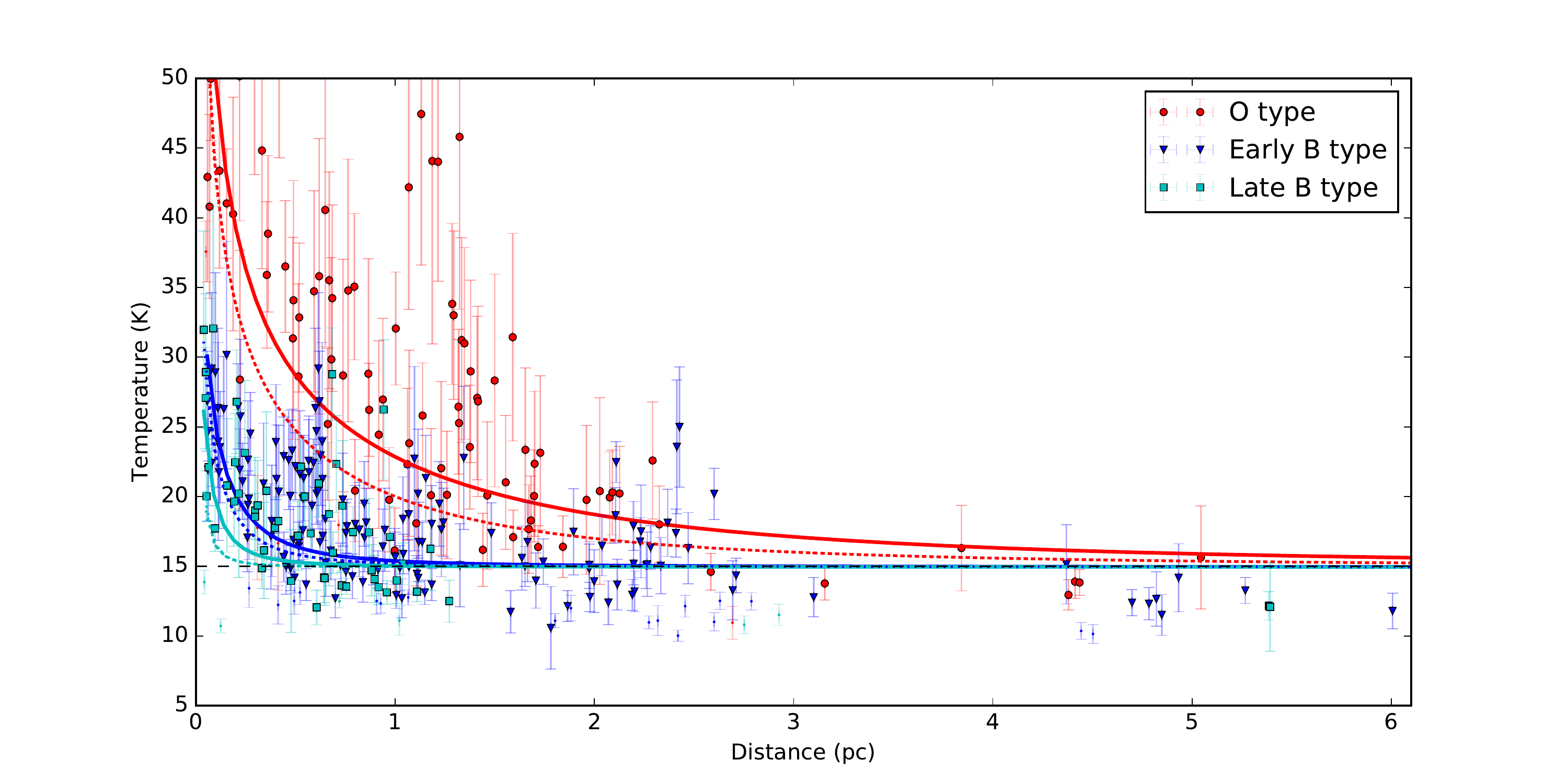}
\caption{Clump temperature as function of projected distance from the major OB~star in the region (red circles: clumps in regions with O stars; blue triangles: early-type B stars; cyan squares: late-type B stars). The solid lines are OB + ISRF heating fits as described in the text (colours as data).  The dotted lines show the expected temperatures vs. distance for clumps heated by stars at the spectral type boundaries with extinction $A_v = 0.5$ (B0: red dots, B5: blue dots; A0: cyan dots); the heated clumps in each spectral type class should lie above these.  Low temperature points excluded from the fits are shown as small `+' .  See Sect.~\ref{sect:results} for details of models and fit.} 
\label{fig:heating}
\end{centering}
\end{figure*}

The variation of clump temperature with projected distance from the region's major OB star is shown in Figure~\ref{fig:heating}.  As expected, more of the clumps in regions with O~stars are warm (20--50~K), consistent with the temperature distributions in Figure~\ref{fig:histograms}.  All subsets show the anticorrelation of clump temperature with projected distance identified in the Aquila W40 complex \citep{Rumble:2016kx}.   Nevertheless, there is some scatter around this trend: projected distances that are shorter than distances along the line-of-sight will produce clumps that are colder than expected and clumps that are locally heated by stars other than the major regional ones listed in Table~\ref{tbl:regions} will be warmer. 

Following \citet{drainebook}, we model the variation in dust temperature  by considering a dust parcel of mass $M$ that is heated by the incident power from an OB star, $\Gamma_\mathrm{OB}$,  plus the interstellar radiation field (ISRF), $\Gamma_\mathrm{ISRF}$, and cooled by the power radiated by the dust, $\Lambda_\mathrm{cool}$.  In thermal equilibrium,

\begin{equation}
  \Lambda_\mathrm{cool} = \Gamma_\mathrm{OB} + \Gamma_\mathrm{ISRF}\label{eqn:cooling}.
\end{equation}

 The OB~star heating varies as the inverse square of distance $d$ and consists of the integrated flux absorbed by the parcel over all wavelengths, 

\begin{equation}
  \Gamma_\mathrm{OB} = {1\over 4\pi d^2} \int L_*(\nu)e^{-\tau_\nu(A_v)}\kappa(\nu) M d\nu,
  \label{eqn:OBheating}
  \end{equation}
where $L_*(\nu)$ is the OB star luminosity, $\kappa_\nu$ is the dust emissivity responsible for the absorption and $M$ the mass.  A layer of extinction between the star and clump is factored as $ e^{-\tau_\nu(A_v)}$ where $\tau_\nu(A_v)$ is the optical depth as a function of visual extinction $A_v$.

Additionally, an incident interstellar radiation field (ISRF) with mean intensity $J_\mathrm{ISRF}(\nu)$ heats the dust mass from all directions.  Assuming no local attenuation of the ISRF, the absorbed power is given by
\begin{equation}
  \Gamma_\mathrm{ISRF} = 4\pi\int J_\mathrm{ISRF}(\nu)\kappa(\nu) M d\nu
  \end{equation}

The dust cools radiatively in all directions following the Stefan-Boltzmann law,
\begin{equation}\Lambda_\mathrm{cool} = 4 \langle\kappa_\nu\rangle M \sigma T_d^4,
  \label{eqn:lambda_cool}
  \end{equation}
  where $T_d$ is the dust temperature  and $4 \langle\kappa_\nu\rangle M$ is the effective surface area of the dust mass.  $\langle\kappa_\nu\rangle$ is the emission-averaged opacity weighted by a blackbody at temperature $T_d$.


In this \begin{jh}formulation\end{jh}, the mass $M$ appears on both sides of the equation and cancels out.  The integrals are carried out over a range in frequency from the equivalent of centimetre wavelengths (for which dust opacity is negligible) to the Lyman limit (91.2~nm), assuming all the shorter-wavelength UV emission is absorbed by hydrogen atoms.

We assume \citet{Ossenkopf:1994vn} model~5 dust as appropriate for dense molecular clouds.  The long-wavelength spectral index of $\beta=1.8$ gives a Planck-averaged dust opacity that depends on temperature as $\langle\kappa\rangle_ {Td} \simeq 10\hbox{ cm}^2\hbox{ g}^{-1} \times (T_d/{10 \hbox{ K}})^\beta $.  We use the same dust properties for the heating/cooling balance and for the extinction layer. 
For the ISRF intensity spectrum, we use the tabulated version from \cite{Evans:2001kx} which has contributions from the microwave background \citep{Wright:1991aa},  IR/visible/UV \citep{Mathis:1983dq}  and short UV \citep{Draine:1978aa}.  With these choices, the power absorbed from the ISRF and the cooling factor per unit mass are $\Gamma_{\mathrm{ISRF}} = 2.33 \times 10^{-2}\hbox{ W}$ and $\Lambda_{\mathrm{cool}} = CT^{4+\beta}$ with $C = 3.59\times 10^{-9} \hbox{ W K}^{-5.8}$.

The heating/cooling balance gives a temperature of 15~K for clumps only heated by the ISRF.  This value is within the range of steady-state dust temperatures calculated by \citet{Mathis:1983dq}.  It is also consistent with the median temperature of clumps in regions without OB stars (see Fig.~\ref{fig:histograms}), where the emission-weighted dust temperatures are dominated by the ISRF-heated outer layers of clumps.  Gas temperatures are expected to be a few Kelvin lower than dust temperatures in these outer layers as the gas is heated primarily by collisions with the dust \citep{forbrich:2014aa}.  \begin{jh}The strength of the ISRF varies from region to region depending on the local stellar distribution and extinction \citep{Bernard:2010}, and regions with significant numbers of clumps and cloud area below the 15~K median indicate a weaker ISRF (see Fig.~\ref{fig:histograms} and Fig.~\ref{fig:heating}).\end{jh}

\begin{table*}
\caption{OB spectral types, power absorbed per unit mass of dust from ISRF, power absorbed per unit mass of dust from OB star without extinction, power absorbed at 1~pc with extinction $A_v=0.5$, least squares fit power at 1~pc and heating range for dust temperatures above 18~K.  See Sect.~\ref{sect:results} for details of models and fit.}
\label{tbl:tdfit}
\begin{tabular}{l l l l l l}
\hline
\hline
  Type & $\Gamma_\mathrm{ISRF}$ & $\Gamma_\mathrm{OB}$ & $\Gamma_\mathrm{OB}(A_v=0.5)$ & $\Gamma_\mathrm{OB}$ (fit) & range\\
  & W / kg & W / kg& W / kg& W / kg & pc\\
  \hline
  O9.5 -- 08 & $2.33\times 10^{-2}$ & 5.2 -- 9.2                        & $(1.2\hbox{ -- }  1.9) \times 10^{-1}$ & $2.5\pm 0.3 \times 10^{-1}$  & 2.4\\
B4 --B0     & $2.33\times 10^{-2}$ & $(4.3 \hbox{ -- } 450)\times 10^{-2}$ & $(3.0 \hbox{ -- }100) \times 10^{-3}$ & $4.2\pm 0.7 \times 10^{-3}$ &0.4 \\
  A0 -- B5    & $2.33\times 10^{-2}$ & $(0.9 \hbox{ -- } 31)\times 10^{-3}$  & $(1.7\hbox{ -- }24)\times 10^{-4} $   & $8.5 \pm 2.4\times 10^{-4}$ &0.2\\
\hline
\end{tabular}
\end{table*}

We calculate the OB heating power $\Gamma_\mathrm{OB}$ for a range of stellar spectral types and extinctions $A_v$.  Table~\ref{tbl:tdfit} gives the values of $\Gamma_\mathrm{OB}$ for unextincted OB stars and OB stars with an extinction layer $A_v = 0.5$.  Combining the heating from OB stars with that from the ISRF as above, $\Gamma_\mathrm{OB} + \Gamma_\mathrm{ISRF}$, the dust temperature was calculated using Equations~\ref{eqn:cooling} and \ref{eqn:lambda_cool}.   Results for B0, B5 and A0 spectral types with an extinction layer $A_v = 0.5$ are shown for comparison with the data in Figure~\ref{fig:heating}.  Unextincted OB stars produce temperatures that are too high by many Kelvin.
The addition of a layer of extinction $A_v \sim 0.5$ between the OB star and the clump, however, reduces the dust temperature to moderate levels consistent with those observed.  
The impact of this assumption is similar to that of the $A_V = 1$ layer assumed by \citet{Galli:2002} when modelling prestellar cores.  Alternatively, these results suggest that the emission-weighted dust temperatures are typical of dust that is at an extinction of $A_V = 0.5$ within the clumps, consistent with results for isolated dense cores such as CB68 \citep{Nielbock:2012,Lippok:2016}.  This subtlety could be explored further with a more sophisticated radiative transfer model, though a diverse range of structures would have to be considered, as the clumps here are not specifically designed to represent individual starless/prestellar/protostellar cores.

\begin{jh}
Projection effects can also reduce the dust temperature: a clump always has a higher actual distance than its projected distance from the OB star, reducing the $1/d^2$ term in the OB~heating (Equation~\ref{eqn:OBheating}).  Projection effects are most significant for clumps at small projected distance but large line-of-sight distance from the OB star, such as the front/back of shells, and are unimportant if the cloud is filamentary in the plane-of-the-sky and clumps are offset along the filament.  The overall effect is to increase the scatter in temperatures at every distance, and particularly at small projected distance.   As cloud geometry is varied and uncertain, and to estimate even an average correction to the  temperatures requires assumptions about the distribution of clumps, in our modelling, we take account of the projected distance simply by excluding a small fraction of clumps with particularly low temperatures for a given distance (see below).
\end{jh}

Lastly, the dust temperatures produced by the model with combined OB~star and ISRF heating were fitted to the measured dust temperatures as a function of distance (as shown in Figure~\ref{fig:heating}) to determine a characteristic value of the OB heating parameter $\Gamma_\mathrm{OB}$ for each of the spectral type categories.  The fitting was carried out using a nonlinear least-$\chi^2$ fit (MPFIT \footnote{https://code.google.com/archive/p/astrolibpy/source/\\default/source}).  
The fit results for the three stellar spectral type ranges are given in Table~\ref{tbl:tdfit} and overplotted on Figure~\ref{fig:heating}.  We excluded clumps within a JCMT beam radius of the OB star from the fit, because their fluxes are affected by free-free emission  and discs, if present,  and iteratively excluded clumps with temperatures well below the model to avoid fitting those clumps that likely have significantly greater actual distances than their projected distance  (i.e., those that were $> 3\sigma$ below the fit after 10 iterations), ultimately excluding 9\% of clumps.

\section{Discussion and conclusions}

The amount of heating and range of the heating varies significantly with the spectral type of the major OB star in the region.  Taking the range of influence of the stars as the distance at which the temperature drops below \begin{jh}18~K\end{jh}, the O-type stars can be seen to have an impact over more than 2~parsecs, essentially the entire region in which they appear, whereas the B stars' influence is less than 0.5~pc (0.4~pc for early and 0.2~pc for late-type stars).  
The impact of O~stars on dust heating is large, with clump dust temperatures raised to 40~K close to the OB star and above 20~K to the furthest extent that clumps are detected in the map (for example, the 2.5~pc transverse extent of the W40 and Orion~B NGC2024 maps).  \begin{jh}Although the absolute values of these temperatures and thresholds will vary if dust opacities vary (where we have assumed a fixed $\beta$) the trend is robust.\end{jh}  The impact of the heating is also apparent as bright emission in the mid-infrared (e.g. \emph{Herschel}  70~\micron\ in Aquila; \citealt{Bontemps:2010fk} or \emph{Spitzer} 24\micron\ in Orion~B; \citealt{Megeath:2012} ).  Mid-infrared emission can be produced by small masses of hot dust (above around 50~K) whereas the submillimetre-based temperatures suggest that the heating also penetrates deeper into the clumps to at least $A_v=0.5$.  The O-star heating range is larger than the radio \HII\ region extent ($6' \sim 0.8$~pc for W40 \citealt{Rodney:2008ij,Vallee:1991zr}, $0.98$~pc for NGC2024 \citealt{Gordon:1969}), which indicates that the penetration of the O-star radiation varies with direction, consistent with a clumpy medium, and supported by the larger size of the associated infrared nebulosity than the radio \HII\ region.

By contrast, the heating range of B-type stars is typically less than half a parsec.  At the luminous end of the range, the power absorbed by dust at 1~pc from a B0 star (with extinction of $A_v = 0.5$) is increased by a factor of 5.3 over that from the interstellar radiation field and results in a factor $1.3$ increase in the dust temperature (as  $\Gamma_\mathrm{cool} \propto T_d^{4+\beta}$), raising temperatures from 15~K to 20~K (see red dashed line on Fig.~\ref{fig:heating})), but this reduces rapidly with spectral type.  For late B~types, the heating range of 0.2~pc is comparable with the typical 0.1~pc size of a protostellar core, suggesting that any radiative feedback will be limited to only to the immediate environment. This range is consistent with the location of heated clumps observed in Serpens MWC~297 \citep{Rumble:2015vn}, Perseus~NGC1333 (\citealt{Hatchell:2005fk}, \citealt{Chun-Yuan-Chen:2016kx}) and L~1495 \citep{Buckle:2015vn}, for example.

Splitting the sample into 179 `heated' and 564 `unheated' clumps (those that lie within the heating ranges listed in Table~\ref{tbl:tdfit}), the `heated' clumps have a mean temperature of 24.0~K, 
significantly larger than 15.5~K for the `unheated' clumps.  


Is stellar radiative heating influencing the subsequent star formation?  We expect the main effect of heating on core stability is to increase the thermal support in the inner, densest regions of cores ($n > 10^{4.5}\hbox{ cm}^{-3}$;  \citealp{Goldsmith:2001fk}) where the dust and gas are well coupled.  Models of externally-heated cores show that gas temperatures interior to the core rise by a similar factor due to FIR emission from small grains penetrating the interior where the dust and gas temperatures are well coupled \citep{Galli:2002}.  This coupling is supported by temperatures measured in the dense gas by the Green Bank Ammonia Survey NH$_3$, which rise above 20~K near the B~stars in NGC1333 (Perseus) and L1688 (Ophiuchus) \citep{Friesen:2017}.  \begin{jh} Although the dust temperatures are generally higher than the gas temperatures (as expected from theory and supporting our choice of $\beta$), the difference falls to less than 2~K towards the dense pre/protostellar cores, with gas and dust kinetic temperatures both approaching 20~K in the dense cores within 0.4~pc of the B2{\sc II} star S1 in Ophiuchus (in NGC1333, there are no dense cores within heating range of the B5~star).  Temperatures of around 20~K for these cores are also found from the joint {\em Herschel}--SCUBA-2 analysis of Ophiuchus which solves for $T$ and $\beta$ independently \citep{Howard:2021}.   In both Ophiuchus and Perseus, gas and dust temperatures rise towards the heated edge of the cloud, with the dust temperature rising more rapidly than the gas temperature as opacities drop and dust cooling reduces.
\end{jh}
With clump temperatures enhanced by 55\% on average, depending on distance from the star, then the thermal support term in the virial equation is increased proportionally.  As most cores lie within a factor two of virial equilibrium, this extra heating could stabilise cores in the `heated' clumps against gravitational collapse.  

  Radiative feedback can also have a substantial effect on the stellar initial mass function (IMF), suppressing fragmentation and reducing the number of low-mass stars  \citep{bate:2009uq, Krumholz:2011, Mathew:2020}.  This impact can result from even moderate temperature rises as the critical mass (Jeans mass $M_J$ or Bonnor-Ebert mass $M_{BE}$) is proportional to $T^{3/2}$.  Namely, the 55\% mean increase in temperature from 15~K to 24~K that we see here (typical of B0 stars at 0.5~pc) raises the critical mass by a factor of two.

A secondary effect of OB stellar heating can be to raise the external pressure in the gas surrounding the clumps.  This pressure is not well traced by the dust temperature partly because the dust and gas temperatures are decoupled in these low density regions but mostly because the main contribution to the pressure comes from turbulence.  External pressure counteracts thermal pressure and acts alongside self-gravity to bind cores.  From a virial analysis in Ophiuchus using linewidths of C$^{18}$O as a tracer, \citet{Pattle:2015ys} found that external pressure is responsible for binding several starless cores, particularly those around the B star S1.  Indeed, the Green Bank Ammonia Survey is finding similar results in other regions \citet{Kerr:2019}.  Further studies are needed to understand if the increase in external pressure counteracts the increase in internal support in the heated regions and mitigates the consequences of stellar heating for the star formation efficiency and IMF.

OB stars can produce much more dramatic changes to star formation regions than simply heating dust.  For example, molecular material can be ionised, photodissociated, collected, disrupted, shocked and moved around.  The overall positive and negative effects of OB star feedback are hard to disentangle \citep{Dale:2015}.  Nonetheless, given that we find that 24\% of clumps are heated externally, even in the nearer star-forming regions with few massive stars, simulations which take into account the effects of radiative feedback, and not just ionising UV radiation, from nearby high- and intermediate-mass stars are needed.

\section*{Data availability}

The data underlying this article are available at \url{https://dx.doi.org/10.11570/20.0007}

(temperature maps and uncertainties in FITS format). The JCMT SCUBA-2 observations are available under Proposal~IDs MJLSG31-41 (PI: Hatchell) from the JCMT archive at the Canadian Astronomy Data Centre \url{https://www.cadc-ccda.hia-iha.nrc-cnrc.gc.ca/en/jcmt/}.   A recent reduction of the GBS 450~\micron\ and 850~\micron\ maps is publicly available at \url{https://doi.org/10.11570/18.0005} \citep{Kirk:2018}.  

\section*{Acknowledgements}

This work is based on observations from JCMT, which has historically been operated by the Joint Astronomy Centre on behalf of the Science and Technology Facilities Council of the United Kingdom, the National Research Council of Canada and the Netherlands Organisation for Scientific Research.  Additional funds for the construction of SCUBA-2 were provided by the Canada Foundation for Innovation.  The authors thank the JCMT staff for their support of the GBS team in data collection and reduction efforts.  They also wish to recognize and acknowledge the very significant cultural role and reverence that the summit of Maunakea has always had within the indigenous Hawaiian community. We are most fortunate to have the opportunity to conduct observations from this mountain.  The data processing was carried out using Starlink software (Currie et al. 2014), which is supported by the East Asian Observatory, and relied on the services of the Canadian Advanced Network for Astronomy Research (CANFAR) which in turn is supported by CANARIE, Compute Canada, University of Victoria, the National Research Council of Canada, and the Canadian Space Agency, and the facilities of the Canadian Astronomy Data Centre operated by the National Research Council of Canada with the support of the Canadian Space Agency.  The data analysis made extensive use of Python modules astropy, matplotlib and APLpy.  D. Rumble was supported by a STFC studentship (ST/1199394) at the University of Exeter and J. Hatchell by the University of Exeter Astrophysics Group.  Our thanks to James di Francesco from the JCMT Gould Belt Survey team who kindly provided constructive comments and suggestions before submission, and to Erik Rosolowsky for his insightful refereeing.


\bibliographystyle{mnras}
\bibliography{OBheating}

\end{document}

%% file: table1.tex
AurigaCentral	&4:25:39	&37:07:01.4	& 470$^{a}$	& 8	& 3	&--	&--	\\
AurigaLkHa101	&4:29:56	&35:15:35.8	& 470$^{a}$	&28	&25	&LKHa101	&B0V$^{1}$	\\
AurigaNorth	&4:09:54	&40:06:28.2	& 470$^{a}$	& 9	& 3	&--	&--	\\
Cepheus L1228	&20:57:13	&77:35:43.8	& 352$^{a}$	& 2	& 1	&--	&--	\\
Cepheus L1251	&22:38:49	&75:11:31.2	& 352$^{a}$	& 8	& 3	&--	&--	\\
Cepheus South	&21:01:31	&68:11:15.3	& 341$^{b}$	&12	& 6	&HD200775	&B2Ve$^{2}$	\\
CrA	&19:01:55	&-36:57:44.5	& 151$^{a}$	&15	&12	&RCrA	&B5III$^{3}$	\\
IC5146	&21:47:23	&47:32:12.0	& 751$^{b}$	&32	&13	&BD+463474	&B0V$^{4}$	\\
Lupus	&15:45:00	&-34:17:07.7	& 151$^{a}$	& 7	& 1	&--	&--	\\
OphSco-L1688HD	&16:26:28	&-24:24:00.2	& 139$^{a}$	&65	&26	&HD147889	&B4Ve$^{5}$	\\
OphSco-L1688S1	&16:26:28	&-24:24:00.2	& 139$^{a}$	&65	&26	&S1	&B2II$^{6}$	\\
OphSco-L1689	&16:32:23	&-24:28:36.5	& 147$^{c}$	&27	& 9	&--	&--	\\
OrionA	&5:35:14	&-5:23:54.5	& 432$^{b}$	&391	&386	&ONC	&O7$^{7}$	\\
OrionB L1622	&5:54:25	&1:44:19.7	& 423$^{a}$	& 5	& 1	&--	&--	\\
OrionB N2023	&5:41:41	&-2:17:09.3	& 423$^{a}$	&27	&24	&HD37903	&B3IV$^{8}$	\\
OrionB N2024	&5:41:45	&-1:55:42.3	& 423$^{a}$	&68	&49	&BCB89IRS2b	&O8$^{1}$	\\
OrionB N2068	&5:47:05	&0:21:45.1	& 423$^{a}$	&83	&58	&HD38563	&B2II+B2III$^{9}$	\\
PerseusIC348	&3:43:57	&32:00:50.9	& 321$^{d}$	&22	&13	&HD281159	&B5V$^{10}$	\\
PerseusNGC1333	&3:29:11	&31:13:33.9	& 294$^{a}$	&36	&35	&LZK12	&B5$^{11}$	\\
PerseusWest	&3:25:36	&30:45:16.3	& 294$^{a}$	&26	&18	&--	&--	\\
Pipe B59	&17:11:22	&-27:26:02.2	& 180$^{b}$	& 5	& 5	&--	&--	\\
SerpSouth	&18:30:04	&-2:03:05.1	& 484$^{a}$	&45	&40	&--	&--	\\
SerpensE	&18:37:51	&-1:45:48.9	& 484$^{a}$	&36	&20	&--	&--	\\
SerpensMWC297	&18:27:59	&-3:49:46.3	& 383$^{e}$	& 9	& 0	&MWC297	&B1.5Ve$^{12}$	\\
SerpensN	&18:39:11	&0:32:50.6	& 484$^{a}$	&13	& 7	&--	&--	\\
Serpens Main	&18:29:50	&1:15:18.9	& 436$^{f}$	&17	&17	&HD170634	&B8$^{13}$	\\
Taurus L1495	&4:18:40	&28:23:18.0	& 141$^{a}$	&10	& 2	&V892Tau	&A0Ve$^{12}$	\\
Taurus South	&4:35:37	&24:09:19.7	& 141$^{a}$	& 5	& 1	&--	&--	\\
Taurus TMC	&4:39:54	&26:03:09.2	& 141$^{a}$	& 2	& 1	&--	&--	\\
W40	&18:31:21	&-2:06:20.3	& 484$^{a}$	&56	&47	&OS1aS	&09.5V$^{14}$	\\